\documentclass[12pt,referee]{aastex61}
\received{2018 January 10}
\accepted{2018 March 02}
\submitjournal{ApJ}

\shorttitle{Observations of surface and body modes in pores}
\shortauthors{P.H. Keys et al.}

\begin{document}

\title{Photospheric observations of surface and body modes in solar magnetic pores}

\author{Peter H. Keys}\affiliation{Astrophysics Research Centre, School of Mathematics and Physics, Queen's University Belfast, Belfast, BT7~1NN, Northern Ireland, U.K.}
\affiliation{Solar Physics and Space Plasma Research Centre (SP$^2$RC), University of Sheffield, Hicks Building, Hounsfield Road, Sheffield, S3~7RH, U.K.}
\author{Richard J. Morton}\affiliation{Mathematics and Information Sciences, Northumbria University, Newcastle Upon Tyne, NE1~8ST, U.K.}
\author{David B. Jess}\affiliation{Astrophysics Research Centre, School of Mathematics and Physics, Queen's University, Belfast, BT7~1NN, Northern Ireland, U.K.}\affiliation{Department of Physics and Astronomy, California State University Northridge, Northridge, CA 91330, U.S.A}
\author{Gary Verth}\affiliation{Solar Physics and Space Plasma Research Centre (SP$^2$RC), University of Sheffield, Hicks Building, Hounsfield Road, Sheffield, S3~7RH, U.K.}
\author{Samuel D. T. Grant}\affiliation{Astrophysics Research Centre, School of Mathematics and Physics, Queen's University, Belfast, BT7~1NN, Northern Ireland, U.K.}
\author{Mihalis Mathioudakis}\affiliation{Astrophysics Research Centre, School of Mathematics and Physics, Queen's University, Belfast, BT7~1NN, Northern Ireland, U.K.}
\author{Duncan H. Mackay}\affiliation{School of Mathematics and Statistics, University of St Andrews, St Andrews, KY16~9SS, U.K.} 
\author{John G. Doyle}\affiliation{Armagh Observatory \& Planetarium, College Hill, Armagh, BT61~9DG, U.K.}
\author{Damian J. Christian}\affiliation{Department of Physics and Astronomy, California State University Northridge, Northridge, CA 91330, U.S.A}
\author{Francis P. Keenan}\affiliation{Astrophysics Research Centre, School of Mathematics and Physics, Queen's University, Belfast, BT7~1NN, 
Northern Ireland, U.K.}
\author{Robertus Erd{\'{e}}lyi}\affiliation{Solar Physics and Space Plasma Research Centre (SP$^2$RC), University of Sheffield, Hicks Building, Hounsfield Road, Sheffield, S3~7RH, U.K.}
\affiliation{Debrecen Heliophysical Observatory (DHO), Research Centre for Astronomy and Earth Sciences, Hungarian Academy of Sciences, 4010 Debrecen, P.O. Box 30, Hungary}

\correspondingauthor{Peter H. Keys}
\email{p.keys@qub.ac.uk}


\begin{abstract}
Over the past number of years, great strides have been made in identifying the various low-order magnetohydrodynamic 
wave modes observable in a number of magnetic structures found within the solar atmosphere. However, one aspect of these 
modes that has remained elusive, until now, is their designation as either surface or body modes. This property has 
significant implications on how these modes transfer energy from the waveguide to the surrounding plasma. Here, for 
the first time to our knowledge, we present conclusive, direct evidence of these wave characteristics in numerous pores which were observed 
to support sausage modes. As well as outlining methods to detect these modes in observations, we make estimates of 
the energies associated with each mode. We find surface modes more frequently in the data, and also that surface modes 
appear to carry more energy than those displaying signatures of body modes. We find frequencies in the range of 
$\sim$2 to 12~mHz with body modes as high as 11~mHz, but we do not find surface modes above 10~mHz. It is expected that the 
techniques we have applied will help researchers search for surface and body signatures in other modes and in differing structures to those presented here. 
\end{abstract}


				
\section{Introduction}
\label{Intro}
The solar atmosphere is a highly dynamic magnetised plasma, whose structure is largely determined by the complex magnetic field that permeates through the layers. This gives rise to many of the features and phenomena frequently observed in the solar atmosphere. The advent of improved instrumentation and techniques has allowed many of their properties to be rigorously studied in recent years. 

Undoubtedly, one of the most interesting aspects associated with the Sun's magnetic field, and which are frequently studied, are magnetohydrodynamic (MHD) wave phenomena. At their most basic, there are three possible MHD wave modes: the incompressible Alfv{\'{e}}n wave, and the slow and fast magnetoacoustic waves \citep{GoedbloedPoedts2004, NakariakovVerwichte2005}. Various wave modes have been observed across numerous features in the different layers of the solar atmosphere \citep[see reviews by][to name a few]{Banerjee2007, Wang2011, DeMoortelNakariakov2012, Mathioudakis2013, Jess2015}. The Sun's convection zone excites a wide spectrum of global acoustic waves ($p$-modes), and when these interact with magnetic flux tubes embedded in the photosphere they excite MHD wave modes with dominant periods of around 300~s \citep{Braun1988, Sakurai1991}. This MHD wave energy is then guided by the flux tubes to higher atmospheric layers. It is still not clear what happens to the upward propagating wave energy, but there is evidence that $p$-modes play a pivotal role in governing the dynamics of the chromosphere, with shocks launching, e.g., chromospheric jets \citep{DePontieu2004}. Furthermore, there are clear indications that the wave energy reaches the corona as e.g. slow magnetoacoustic waves \citep{DePontieu2005, DeMoortel2009}, or is transferred to transverse motions that could potentially play a role in heating the coronal volume \citep[to name a few]{Morton2012, Freij2014, Grant2015, Morton2015}.

The foundation for the theoretical description of MHD waves in solar magnetic waveguide models, as it is widely used today, was formulated in the early 1980's in seminal papers by, e.g., \citet{Spruit1982} and \citet{EdwinRoberts1983}. Driven by observations, internal and external background quantities such as plasma density and magnetic field strength are allowed to vary, resulting in magnetic waveguides capable of supporting a much richer variety of MHD modes than are present in a homogeneous infinite plasma. This is most clearly seen from the dispersion diagrams of such waveguides, which display a complex variety of weakly and strongly dispersive magnetoacoustic wave modes, depending on, e.g., waveguide width, wavenumber, plasma beta, and internal/external Alfv{\'e}n and sound speeds \citep{EdwinRoberts1983}. Also, the spatial structure of these wave modes is fundamentally determined by the cross-sectional shape of the waveguide. For example, a flux tube with circular cross-section supports, e.g., azimuthally symmetric (sausage), asymmetric (kink) and higher order perturbations (fluting modes). Another key property of such waveguides is whether the wave mode is evanescent in the external plasma, i.e., trapped by the waveguide, or oscillatory outside, i.e., leaky. 

Often pores are employed to study sausage modes. Pores are relatively small ($\sim$1\,--\,6~Mm in diameter) and have field strengths of the order of a kilogauss \citep{Sobotka2003}. Their small size means that they are more dynamic and responsive to external forces. Like sunspots, pores are darker than the quiescent solar surface. One study \citep{VermDen2014} measured the mean intensity of a large sample of pores as being up to 40\% below the surrounding surface, while more recent work \citep{Dorotovic2016} on several pores using satellite data showed that they form when the intensity drops below 0.85 of the surrounding photospheric intensity value and the magnetic field increases to 650~G. Unlike sunspots, however, pores are devoid of penumbrae, meaning they are fairly simple magnetic structures with lifetimes from several hours up to days \citep{Sutterlin1996}.

To observe sausage modes in ground-based data, several studies \citep{Dorotovic2008, Moreels2015a, Grant2015, Freij2016} searched for oscillatory signals in the cross-sectional area and intensity of pores. A key conclusion in each of these works was that the fractional variations in both area and intensity was so minor, that ground-based data was essential for studies of sausage modes in lower atmospheric regions. The first investigation of sausage modes \citep{Aschwanden2004}, observed them in large-scale loop oscillations in the corona, while the initial evidence of sausage modes lower in the atmosphere arrived several years later \citep{Dorotovic2008} by analysing white-light channels with the Swedish Solar Telescope (SST). However, the latter authors only searched for oscillatory signals in cross-sectional area and not concurrent intensity oscillations. Subsequent work by \citet{Morton2011} conducted a more thorough investigation, looking for signals in both area and intensity oscillations in a blue continuum (4170{\AA}) channel, finding periods from 30~s to 450~s. These authors noted that these periods would suggest that excitation of the sausage modes was due to global $p$-mode oscillations, and they also stated that the modes did not have large amplitude wave power, provided that there were no twists in the magnetic field. A study of phase relationships between the area and intensity signals by \citet{Moreels2013} highlights that these signals are always in-phase for slow modes, while they are in anti-phase for fast modes. \citet{Moreels2013} suggest that \citet{Morton2011} observe the fast sausage mode. \citet{Dorotovic2014} studied both pores and sunspots for signatures of sausage modes, and report both slow and fast modes, with periods ranging from 4 to 65~minutes. These results indicate that sausage modes can be excited in a range of photospheric structures of varying size and shape.

Observations of sausage modes in the chromosphere followed their discovery in the photosphere \citep{Morton2012}. Here the authors find kink and sausage modes in chromospheric fibrils, estimate associated energies of 11\,000~W\,m$^{-2}$, and also that the modes were leaky. Therefore, it is possible for these modes to dissipate energy in the corona. \citet{Grant2015} employed multiple passbands and instruments to observe an upwardly propagating sausage mode from the lower photosphere to the upper photosphere/low chromosphere. They used the energy equations of \citet{Moreels2015b} to determine that the energy carried by the modes decreases substantially with height and, thus, may release significant energy into the surrounding chromospheric plasma. A recent study of two pores \citep{Freij2016} using magneto-seismology techniques suggested that sausage modes in pores can be standing harmonics, with strong reflection at the transition region, indicating a chromospheric resonator.

One aspect of magnetoacoustic modes that has been predicted in theoretical work \citep{EdwinRoberts1983}, yet has been neglected in observational studies, until now, is the wave character of the modes, i.e, whether they can be classified as surface or body modes. Expressed simply, a surface or a body mode can form at an interface where physical properties vary sharply, and are analogous to seismic waves associated with earthquakes that occur at many tectonic plate boundaries on Earth. As properties such as magnetic field and density vary rapidly from a pore to its surroundings, pores should support surface and body modes. Recent theoretical work \citep{Yu2017a, Yu2017b} showed that resonant damping of slow surface sausage modes could be efficient under conditions usually observed in pores. Some previous studies of pore oscillations \citep{Moreels2015a, Grant2015, Freij2016} infer the surface/body characteristics of sausage modes using a combination of semi-empirical models, theory and the derived parameters of the modes from observations. Understanding the surface and body properties, however, is crucial in determining how energy is dissipated by modes in higher regions \citep{Yu2017b}. Here, for the first time to our knowledge, we present conclusive, direct evidence for the existence of surface and body modes in photospheric pores supporting sausage modes.

\section{Theory applicable to sausage modes}
\label{Theory}
Most of the theory applicable to waves observed in solar pores has been derived previously \citep{Spruit1982, EdwinRoberts1983} employing, e.g., thin flux tube approximations. In our case, the thin tube approximation is not applicable and we must take the finite tube equations into consideration. We do not consider the effects of gravity on propagation here, though it may be important for pores found in the photosphere.

MHD modes which can propagate in a flux tube, such as a pore, under photospheric conditions fall into three distinct bands in terms of phase speed: fast surface modes, slow body modes and slow surface modes. The phase speed for the slow modes is defined by the tube speed ($c_T$), and since the wave will be barely dispersive, is given by,
\begin{equation}
c_{T} = \frac{c_sv_A}{{(c^{2}_{s} + v^{2}_{A})}^{1/2}},
\end{equation}
where $c_s$ is the sound speed and $v_A$ is the Alfv{\'e}n speed. The sound and Alfv{\'e}n speeds are defined as,
\begin{eqnarray}
c_s = \sqrt{\frac{\gamma R T}{\mu}}\\
v_A = \frac{B_z}{\sqrt{\mu_0 \rho}}
\end{eqnarray}
respectively, where $\gamma$ is the ratio of specific heats, R is the gas constant, T the temperature in the pore, $\mu$ the mean molecular weight, $B_z$ the magnetic field component in the $z$ direction, $\mu_0$ the magnetic permeability and $\rho$ the local plasma density.

Slow modes can be further divided into various angular modes, where $m = 0$ denotes the axisymmetric sausage mode in a cylindrical flux tube. The linear theory for sausage modes in a gravitationally-stratified atmosphere has previously been studied in a rigorous manner \citep{Defouw1976, RobertsWebb1978, DiazRoberts2006, LunaCardoza2012}.

Consider a cylindrical waveguide with vertical background magnetic field denoted as ${\mathbf{B_0}} = B_0{\mathbf{\hat{z}}}$ and a velocity perturbation given by ${\mathbf{v_1}} = (v_r,v_{\theta},v_z)$. In the case of sausage modes, where $m = 0$, the equations for $v_r$ and $v_z$ decouple from the governing equation of $v_{\theta}$. Hence, the magneto-acoustic modes will be described by $v_r$ and $v_z$, while the Alfv{\'e}n mode is given by $v_{\theta}$. We are only interested in the magneto-acoustic mode, and we can therefore ignore the $v_{\theta}$ component and also that of the magnetic field in the $\theta$ direction.

Surface and body modes are characterised by the spatial distribution of the amplitude across the flux tube \citep{RaeRoberts1983, Zhugzhda2000, ErdelyiFedun2010}. The maximum amplitude for the surface mode will always occur at the boundary of the flux tube at the sharp, discontinuity between the varying physical parameters of the equilibrium. For the body mode, the position of the maximum amplitude is dependent upon the mode, i.e., the number of nodes in the {\textit{radial}} direction, and the perturbed quantity chosen.

Equations for the amplitude of the internal plasma parameters for the sausage body mode \citep[following][]{Spruit1982} are,
\begin{eqnarray}
v_z \propto J_0 (m kr),\\
v_r \propto \frac{{\mathrm{d}J_0 (m kr)}}{{\mathrm{d}} r},\\
b_z \propto v_z, \\
b_r \propto v_r, \\
p_1 \propto v_z,
\end{eqnarray}
where the Fourier analysed perturbations are assumed to have the form $f_1 \sim \mathrm{exp}(i(kz - \omega t))$. In the equations above, $p_1$ is the perturbation in the kinetic gas pressure and $b$ the perturbed magnetic field. Also, $J_0$ is the Bessel function of zeroth order and 
\begin{equation}
\label{msquared}
m^{2} = \frac{(v^{2}_{A} - c^{2}_{ph}) (c^{2}_{s} - c^{2}_{ph})}{(v^{2}_{A} + c^{2}_{ph}) (c^{2}_{T} - c^{2}_{ph})}.
\end{equation}
Here, $c_{ph}$ is the phase speed of the mode. Note that these relations do not show the phase relations between the different variables under consideration. Taking into account these equations, we can conclude that $v_z$, $b_z$ and $p_1$ have maximum amplitudes at the center of the flux tube, while $v_r$ and $b_r$ are found to have maximum amplitude at the tube boundary. For higher harmonics in the radial direction, there may be nodes between the axis of symmetry and the boundary of the flux tube.

\begin{figure*}[h!]
\makebox[\linewidth]{
   \includegraphics[width=0.75\linewidth]{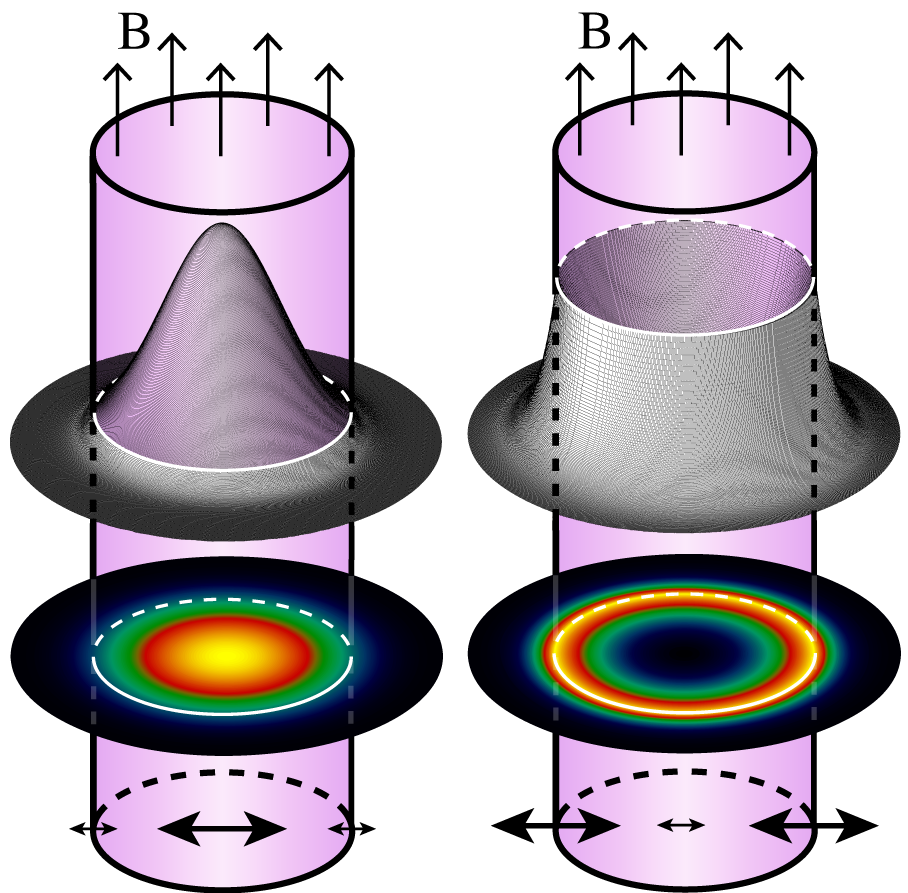}
	}
     \caption{A simplified representation of magnetic flux tubes is shown with the arrows at the top indicating the magnetic field, $\mathbf{B}$. Plasma parameters (e.g., magnetic field, density) of the internal and external plasma differ. Magnetic flux tubes that support the MHD sausage mode are subject to a periodic variation in pressure and area (with these oscillations depicted by the arrows at the bottom). The surface plots (upper images) demonstrate the spatial structure of the pressure perturbation amplitude, which can have two distinct distributions. The amplitude of the body mode (left) is maximal at the central, inner part of the flux tube with the power decaying close to the boundary. On the other hand, surface modes (right) are maximal at the tube boundary defined by the sharp changes in equilibrium quantities modeled as a discontinuity.  The two-dimensional projection of the power is also demonstrated by the colored disks, and can be compared to the observed distributions.
}
     \label{Fig1}
\end{figure*}

In the case of surface modes, the Bessel functions are replaced by the modified Bessel function, $I_0 (n_{0}kr)$, where $-n^{2}_{0} = m^{2} < 0$. Here, {\textit{all}} perturbations have a maximum at the tube boundary and are zero at the center of the tube. This behavior is demonstrated in the schematic shown in Figure~\ref{Fig1}, and shows how one would expect the power plots to look under the ideal scenario for both the body and the surface sausage modes{\footnote{Further visualisations of sausage modes in various simple geometires can be seen here: \url{http://swat.group.shef.ac.uk/fluxtube.html}}}.

\subsection{Predicting surface/body modes}
\label{predicting}

The value $m^{2}$ defined by Equation~\ref{msquared} in the previous subsection can be used to predict semi-empirically if we observe a surface or body mode, depending on whether it is positive or negative in value. To make this estimate we use a combination of the observed physical properties available in our datasets, with some estimates for other parameters from models \citep{VAL1981, Maltby1986} to estimate $m^{2}$ for each oscillation found in each dataset. 

It is clear from Equation~\ref{msquared} that to calculate $m^{2}$ we first must estimate the phase speed, $c_{ph}$. Here, we only employ one bandpass in our analysis. Therefore, we utilise a technique demonstrated previously \citep{Grant2015, Moreels2015b} to yield accurate estimates of the phase speed of a sausage mode oscillation using the equation,
\begin{equation}
c_{ph} = c_s \sqrt{\frac{\pm A_{m} - 1}{\pm A_{m} - 1 + (\gamma - 1) \left( \frac{h \nu}{k_{B} T} \right)}},
\end{equation}
where $h$ is the Planck constant, $\nu$ the frequency of the filter used in our observations, $k_{B}$ the Boltzmann constant and $A_m$ the dimensionless amplitude. Note that the $\pm$ in the formula arises as this expression originates for a quadratic equation. The dimensionless amplitude, $A_m$, can be defined as, 
\begin{equation}
A_m = \frac{^{\delta I}/_{I_0}}{^{\delta A}/_{A_0}}.
\end{equation}
Here, $\delta I$ is the amplitude of the intensity perturbation, $I_0$ the mean intensity, $\delta A$ the amplitude of the area perturbation and $A_0$ the mean area; all of which can be obtained from our observations. This approach of predicting whether a sausage mode is a surface or body mode has been performed previously \citep{Moreels2015a, Grant2015}. Now, we determine how accurate the method is with respect to our direct detection methods.

\section{Observational Data}
\label{obs}
In total we employed 7 datasets of various pores at disc centre from 2011 to 2014. All were acquired with the {\textit{Rapid Oscillations in the Solar Atmosphere}} \citep[ROSA]{Jess2010} instrument. ROSA is a multi-channel broad-band imager installed as a common-user instrument at the Dunn Solar Telescope (DST), New Mexico. 

In this study, we employed the G-band continuum filter centered at 4305.5~{\AA} with a bandpass of 9.2~{\AA} with ROSA. This filter allows us to obtain photospheric intensity images at an estimated continuum formation height of $\sim$100~km \citep{Jess2012} and a theoretical 2-pixel, diffraction-limited resolution of 0$\arcsec$.14 ($\sim$102~km) at a frame rate of 30.3~s$^{-1}$. 

Localised seeing-induced wavefront deformations are corrected in the data in-situ with the use of high-order adaptive optics systems \citep{Rimmele2004}. However, this does not completely correct image deformations and, as a result, we must employ post-facto image reconstruction techniques, such as the KISIP speckle interferometry package \citep{Woger2008} to obtain science-ready images. By utilising 64$\rightarrow$1 restorations, our reconstructed image cadence was increased from 0.033~s to 2.112~s.

In tandem with ROSA, we employ line-of-sight (LOS) magnetograms for all datasets using observations from the {\textit{Helioseismic Magnetic Imager}} \citep[HMI]{Shou2012} instrument onboard the {\textit{Solar Dynamics Observatory}} \citep[SDO]{Pesnell2012}. This data ensures we have magnetic field information to establish estimates for values described in Section~\ref{Theory}. Subsequently, the ROSA and HMI images needed to be aligned. To do this we acquired both the HMI continuum images and the magnetograms for the corresponding times of the ROSA observations, and prepared the data using the standard `{\textit{hmi\_prep}}' routine supplied by the SDO science team. For data alignment, we took a sub-field of the HMI continuum images that represents the telescope pointing for each dataset and degraded the ROSA images to match the resolution of HMI. We then employed Fourier cross-correlation techniques between the HMI continuum sub-fields and the degraded ROSA images to obtain accurate co-alignments. This process resulted in sub-pixel co-alignment between the HMI images and the degraded ROSA images, with maximum $x-$ and $y-$displacements less than one tenth of a HMI pixel. Following such an accurate co-alignment, we could then construct the corresponding sub-fields for the HMI magnetograms for each dataset. Note that the subsequent data analysis was performed on the non-degraded ROSA images.

In selecting data for analysis from the ROSA archive, we impose several criteria:
\begin{enumerate}
\item The data duration is greater than 20 minutes. This ensured there is adequate sampling of any waves present, i.e., the dataset is at least four times longer than the  5~minute frequency often associated with the $p$-mode spectrum.
\item The 2-pixel spatial resolution is better than 0$\arcsec$.5. Due to the fractional area variations associated with sausage modes, we required good spatial sampling of pores. We estimate the spatial resolution using techniques described in \citet{Beck2007}.
\item The cadence of the reduced data had to be shorter than 5~s. This ensures adequate temporal resolution in analysing pore boundary variations, and in determining subsequent oscillations with both wavelet analysis and empirical mode decomposition.
\item The pore data are close to disk center. This removes any LOS effects on boundary/intensity estimates induced by studying pores significantly away from disk center.
\item The pore datasets were acquired within the operational time frame of the SDO, i.e., after February 2010. This ensures that there is adequate magnetic field information from the HMI instrument onboard SDO.  
\item The quality of the datasets, with regard to seeing conditions, is consistent. This ensures a more accurate determination of area and intensity oscillations. If the data contained patches of poor seeing, the resultant area and intensity measurements would result in inaccurate estimations of oscillation periods.
\item We limited the study to simple pores, i.e., we ignored pores with any developing penumbral elements or pores with more complex structuring, e.g., pores separated by light-bridges. By excluding pores with any penumbral elements, there will be less inclination in the magnetic field of the pore. This could possibly affect the analysis of the spatial structuring of the power within the pore, and, as such, affect our determination of surface or body modes in the pore. We employed the HMI images to help isolate less complex pores by using the images to determine structures within the same flux concentration. This was used to determine if neighboring pores are actually separate entities, or if a more complex structure (such as a light bridge) exists within the data. By ignoring pores with more complex structures, one removes complex structures which may exhibit complex oscillatory phenomena, e.g. higher-order standing modes or mixing of modes. If this were the case, then the determination of the spatial structuring of the power would be more difficult, which is essential in determining whether a surface or body mode is observed.
\end{enumerate}

These criteria isolated the 7 datasets we study in this paper taken between 2011 and 2014. Table~\ref{table1} summarises the main characteristics of all datasets used in this study. An example image of each dataset is shown in Figure~\ref{FigA1}.

\begin{table}[hb!]
\centering                                    
\caption{Summary of observations for the sample of pores studied.}           
\label{table1}      
\begin{tabular}{l c c c c}         
\hline\hline                        
\textbf{Data Set} &  \textbf{Spatial Sampling} &  \textbf{Resolution} & \textbf{Pointing} & \textbf{Sequence Duration (mins)}\\
\hline                                   
2011 Jul 11 & 0.$\arcsec$069 & 0.$\arcsec$16 & S16.3, E03.3 & 115 \\
2011 Dec 09 & 0.$\arcsec$069 & 0.$\arcsec$17 & N08.8, E10.0 & 75 \\
2011 Dec 10 & 0.$\arcsec$069 & 0.$\arcsec$21 & N07.6, W04.2 & 118 \\
2012 Sept 30 & 0.$\arcsec$0935 & 0.$\arcsec$20 & S06.7, E00.4 & 30 \\
2013 Mar 06 & 0.$\arcsec$138 & 0.$\arcsec$30 & S17.3, W07.0 & 35 \\
2013 Aug 17 & 0.$\arcsec$069 & 0.$\arcsec$16 & N17.5, E08.6 & 46 \\
2014 Apr 15 & 0.$\arcsec$069 & 0.$\arcsec$17 & S08.9, E04.6 & 51 \\
\hline                                            
\end{tabular}
\end{table}


\section{Analysis and Results}
\label{analy}

\subsection{Wave signatures and mode identification}
\label{FindOscs}
To determine if the signatures of sausage modes are present in our data, a time series of the area and intensity signals for the pores in each dataset is calculated. We define the pore boundary as being any pixel 3$\sigma$ below the mean value of intensity, which is calculated frame-to-frame utilizing a quiet region of the field-of-view (FoV) devoid of network magnetic bright points (MBPs) or the influence of the pore. The boundary is calculated within a box containing the pore under investigation to remove the possibility of counting other pores within the FoV in the area/intensity calculations. Once the pore boundary is established, the area and intensity within the boundary in each frame produces a time series of the variations in these properties over the duration of the dataset for the pore. Concurrently, a time-averaged pore boundary map is created to subsequently identify the pore boundary location when determining the spatial properties of the mode later in the analysis.

Wavelets are employed here as they are considered a standard tool for studying periodic oscillations in signals \citep{TorrCompo1998,Grinstead2004}. In comparison to traditional Fourier methods, where the basis functions are localized only in the frequency domain, wavelet analysis methods are localized in both the frequency and time domains so that the signal is decomposed into both the frequency and time space simultaneously. This allows information to be obtained on both the amplitude of periodic signals and how this amplitude varies over the duration of the signal. In our wavelet analysis we use the 99\% significance level to establish that the periods are real. We use wavelets in the first instance to search for oscillatory signals in both the established area and intensity signals.

Empirical Mode Decomposition (EMD) is a key part of the Hilbert-Huang transform \citep{Huang1998, Terradas2004, Huang2008}, which we employ here as a complementary technique to wavelet analysis. It is a powerful statistical tool which decomposes a signal into its intrinsic timescales. The components are finite in number and are referred to as intrinsic mode functions (IMFs). Decomposition of a signal into its composite IMFs is useful in analyzing both non-stationary and non-linear signals, as the decomposition is based on the local characteristic timescale of the data (i.e., without leaving the time domain). EMD is useful in overcoming some of the limitations of wavelet analysis such as leakage and low time-frequency resolution, which makes it an attractive tool for MHD wave studies.

Each IMF has its own timescale of variation with oscillations symmetric about the local zero mean. As such, the IMF is a function where the number of extrema and zero crossings for each IMF must be either equal or differ, at most, by one. Also, at any point in the IMF, the mean value of the envelope defined by the local maxima and the envelope defined by the local minima must be zero. The first condition reduces period mixing by ensuring that wildly varying periods are not included within the same IMF, while the second condition maintains the local requirement that the oscillations are about zero. 

\begin{figure}[h!]
\makebox[\linewidth]{
   \includegraphics[width=0.8\linewidth]{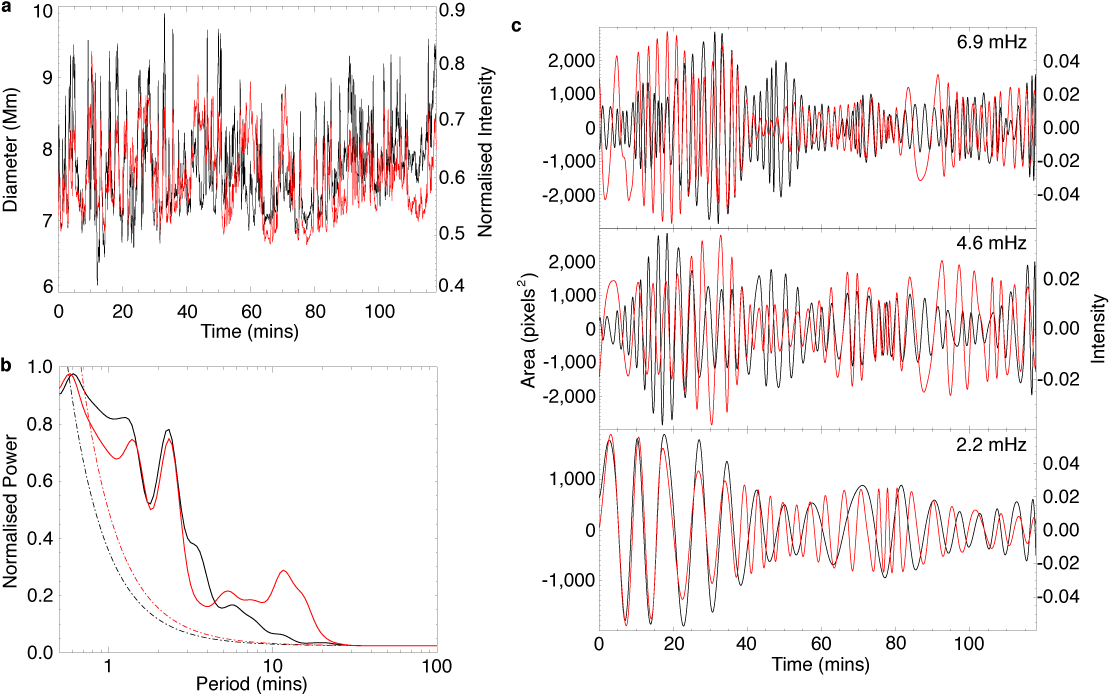}
	}
     \caption{Multiple techniques are employed to determine oscillatory signals present in our data before filtering to observe the spatial distribution of the power. Panel {\textbf{a}} is the area ({\textit{black}}) and intensity ({\textit{red}}) signals established for a single pore over the duration of the observation sequence (2011 December 10 dataset). Panel {\textbf{b}} shows the dominant periods as determined by employing wavelet analysis on both the area and intensity signals with the {\textit{dashed}} lines indicating the 99\% significance levels. Panel {\textbf{c}} shows the result of employing the complementary empirical mode decomposition (EMD) technique on the area and intensity signals in {\textbf{a}}. Intrinsic Mode Functions (IMFs) at three different frequencies for both the area and intensity oscillations are displayed. This information is employed along with wavelet analysis to determine whether a sausage mode oscillation is present in the data, the dominant periods of oscillation and whether the fast/slow sausage mode is observed. 
}
     \label{Fig2}
\end{figure}

Utilizing these time-series and wave identification techniques, the dominant periods within each dataset for both the intensity and area are obtained (Figure~\ref{Fig2}), with both signals displaying significant power with periodicities in the range of 90-700~s. The most common oscillations detected for the two variables, namely intensity and area, have periods $\sim$300$\pm$45~s, consistent with the idea that these waves are excited by the absorption of $p$-modes. The final column of Figure~\ref{FigA1} shows the power as a function of frequency for the pores studied. The frequencies of highest power fall within the range $\sim$2\,--\,5~mHz, which is consistent with the range of freqencies associated with the $p$-mode spectrum \citep[see Fig. 2 of][]{Ludwig2009}. This is further evidence that the wave modes are driven by $p$-modes. However, the detected wave power is not continuous and reveals the waves are composed of coherent wave trains of short duration (Figures~\ref{Fig2}c and \ref{Fig3}). Such behavior has also previously been identified in sunspot waves \citep{BogdanJudge2006}. Corresponding perturbations in the magnetic field in HMI data were not found, likely due to the expected magnetic field variation being of the order of the sensitivity of HMI \citep{Grant2015}.

To determine whether the observed oscillations are slow or fast MHD sausage modes, the phase relationship between the intensity and area signals of the pore is calculated. Due to the presence of discrete packets of oscillatory power, the time-series is evidently not stationary. To accurately assess the cross-spectral phase in the presence of non-stationarity, two methods are employed. The first utilizes wavelets and determines the phase and the coherence by evaluating the cross-spectrum between the two signals, while the second exploits EMD, decomposing the signal into a finite number of IMFs. IMFs of the intensity and area time-series with similar time-scales are then compared for phase relations. These complementary techniques reveal that the intensity is in-phase with the change in area of the pores when a wave packet is identified, suggesting that the observed wave behavior is the compressible slow sausage mode \citep{Moreels2013}. 

Figure~\ref{Fig2} shows the result of these various processes to determine the wave mode, where panel {\textbf{a}} plots the area (\textit{black}) and intensity signals (\textit{red}) for the duration of an observing sequence of a pore (2011 December 10$^{th}$ dataset) obtained from the ROSA archive. Panel {\textbf{b}} displays the dominant periods as determined by employing wavelet analysis on both the area and intensity signals with the {\textit{dashed}} lines indicating the 99\% significance levels, while panel {\textbf{c}} demonstrates the result of employing the complementary EMD technique on the area and intensity signals in {\textbf{a}}. The plots here are the IMFs at three different frequencies for both the area and intensity oscillations for this particular dataset.  Note that, as the waves are composed of coherent wave trains of short duration, the wave power is not continuous. This results in the observed area and intensity signals not being in-phase for the whole duration of the time series as seen in the IMFs in Figure~\ref{Fig2}. Effectively the sausage modes are quasi-periodic as a result of the imperfect waveguides and drivers in the observed data. IMFs as well as wavelet analysis are used to determine the frames that are isolated to observe the spatial distribution of the power for these individual wave trains.

\subsection{Spatial distribution of power}
\label{Spatpow}

The novel aspect of this work is the determination of the spatial structure of oscillatory power. With confidence in the identification of the wave mode, we can now progress with the important objective of determining whether the oscillatory behavior displays the signatures of either the surface or body modes. The power distribution for both the surface and body modes depends on the azimuthal wave-mode number and the perturbed quantity examined. In Figure~\ref{Fig1}, we demonstrate the expected spatial distribution of power of the gas pressure for both the body and surface slow sausage modes (see Section~\ref{Theory} for more details). The kinetic gas pressure amplitude is shown as this represents the variation in the key parameters that govern the image intensity, i.e., gas temperature and density. It is evident that for the sausage body mode, the peak power will generally be concentrated within the center of the waveguide, with the power decreasing towards the pore boundaries. On the other hand, the peak power of the surface sausage mode is located at the boundaries of the wave-guide, decreasing to the center. For both modes, at least in homogenous ambient plasmas, the external wave power should decrease exponentially as a function of distance from the pore boundary. We note that the schematic in Figure~\ref{Fig1} is a visualization sketch, highlighting only the most basic features of surface and body modes. It should be expected (and is observed) that the physical picture is much more complex, with many factors likely contributing to the observed amplitude profile, e.g., radial structuring, variations in cross-sectional geometry and time-dependence due to plasma dynamics. 

\begin{figure}[h!]
\makebox[\linewidth]{
   \includegraphics[width=0.8\linewidth]{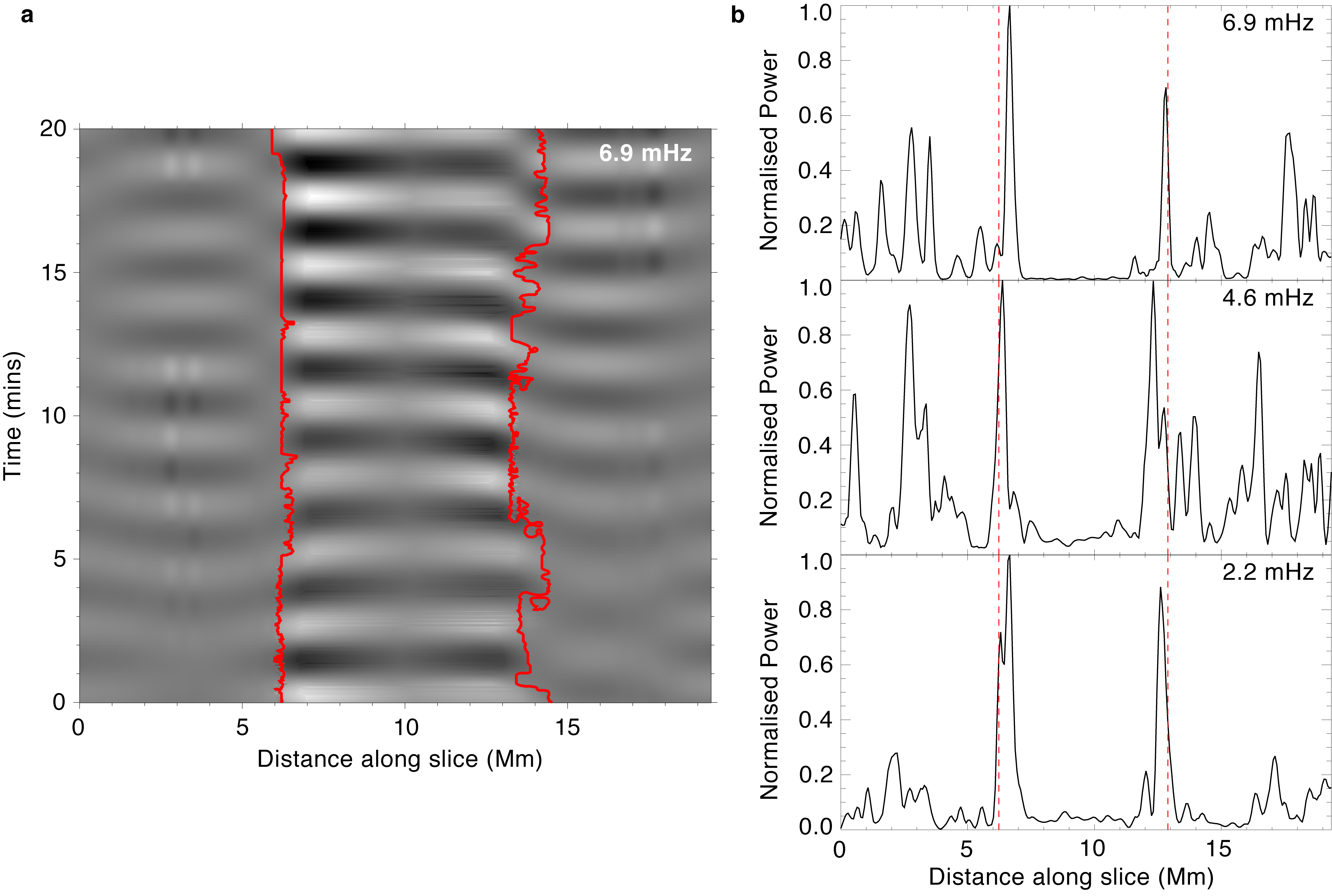}
	}
     \caption{Coherent oscillatory behavior is identified throughout the pore for data from 2011 December 10, and is occurring with a number of distinct frequencies. Panel {\textbf{a}} shows a temporally filtered (at $\sim$6.9~mHz) time-distance diagram taken from a slice across the pore. The analysis reveals a clear, spatially coherent oscillatory pattern within the pore boundaries (outlined in {\textit{red}}). The time axis of the plot is clipped to 20~minutes to ensure that the small pixel-to-pixel amplitude variations across the pore are evident. The time window is established by using EMD and wavelet analysis to determine when a clear in-phase wave is present in the filtered area and intensity signals (Figure~\ref{Fig2}). In the above plot, the 20-minute window equates to the time period of 35 minutes to 55 minutes from the start of data acquisition and represents the time period over which the corresponding plot in panel {\textbf{b}} is derived. The right-hand panels {\textbf{b}} display the normalized Fourier power for three distinct oscillatory frequencies of the pore. The power is seen to peak close to the temporally averaged location of the pore boundary ({\textit{red dashed}} lines), supporting their characterization as a surface mode. Figure~\ref{Fig4} indicates the cut used to make these plots and demonstrates the process of filtering lightcurves at both the pore boundary and pore center for the various oscillations present here.
}
     \label{Fig3}
\end{figure}

To analyze the distribution of power across the pore, the power is examined at selected frequencies for both one-dimensional cuts across the pore and two-dimensional power maps. The power for each frequency examined is averaged over sections of the time-series where significant power was found from the wavelet/EMD analysis. Having determined prominent, periodic perturbations simultaneously between area and intensity signals in the data, the dominant oscillatory frequencies were isolated by employing Gaussian filters to the data. To do this, each pixel within the dataset is treated as a time-varying light curve and converted to frequency space with a Fast Fourier Transform (FFT). This is then convolved with a Gaussian profile with a central frequency, $f$, corresponding to the frequency under investigation, and a width given by $\pm\,f / 10$ to ensure narrow frequency filtering. The filtered data are then converted back into the temporal domain with a FFT. Time-distance plots of the filtered data result in figures such as Figure~\ref{Fig3}a. Using a combination of well-known wavelet and EMD techniques allows us to isolate the frames in the filtered data in which clear, in-phase oscillatory signals in both area and intensity are observed. Focusing on these frames alone allows us to evaluate the 1-dimensional power plots (Figure~\ref{Fig3}b) across a range of angles around the pore, which can be used to determine the 1-dimensional spatial distribution of power across the pore (see also Figure~\ref{FigA1} in the Appendix for more examples).

\begin{figure}[h!]
\makebox[\linewidth]{
   \includegraphics[width=0.8\linewidth]{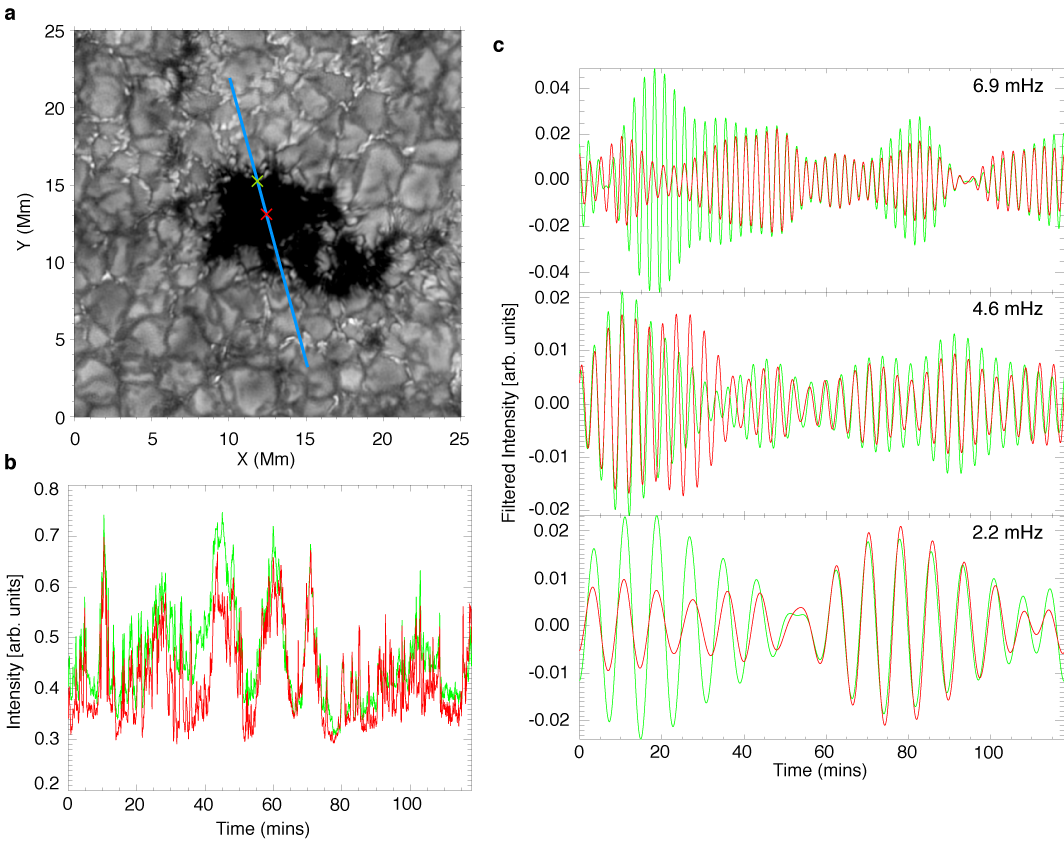}
	}
     \caption{Panel {\textbf{a}} shows a sub-field of the pore under analysis in Figure~\ref{Fig2} with the {\textit{blue}} line indicating the cross-cut used to make the images 
	in Figure~\ref{Fig3}. The {\textit{red}} and {\textit{green}} crosses indicate the locations used to plot the subsequent light curves shown in panel {\textbf{b}} and 
	{\textbf{c}} for a pixel at the center of the pore ({\textit{red}}) and at the pore boundary ({\textit{green}}). Panel {\textbf{b}} shows the unaltered intensity 
	curves for these locations with the color of the line consistent with the crosses in the panel above. Panel {\textbf{c}} shows the results of filtering 
	these light curves as described in the main text over the whole time sequence of the dataset. Again, the colors of the plots are consistent across all panels. 
	It is clear from the filtered plots that there is significant amplitude in intensity oscillations at both the pore boundary and the interior across all three 
	sampled frequencies. The power plots in Figure~\ref{Fig3} panel {\textbf{b}} of the main text are created using frames between $\sim$35\,--\,65~minutes, $\sim$66\,--\,84~minutes and $\sim$1\,--\,29~minutes, from the start of the observing sequence for the 6.9~mHz, 4.6~mHz, and 2.2~mHz oscillations, respectively.
}
     \label{Fig4}
\end{figure}

Figure~\ref{Fig4}a depicts the location of the cross-cut used to create the plots in Figure~\ref{Fig3}. In this panel, a cross along the slice marks the pixel locations used for the subsequent plots displayed in panels {\textbf{b}} and {\textbf{c}}, with one chosen to represent the pore boundary (\textit{green}) and another to represent its centre (\textit{red}). Panel {\textbf{b}} is the collection of untouched lightcurves for these two locations (with the colors used matching the crosses in Panel {\textbf{a}}), while panel {\textbf{c}} is the result of filtering the lightcurves for the three Gaussian filters used for this particular dataset. It is clear from the filtered plots that there is significant amplitude in intensity oscillations at both the pore boundary and the interior across all three sampled frequencies. The power plots in Figure~\ref{Fig3}b are created using frames between $\sim$35~--~65~minutes, $\sim$66~--~84~minutes and $\sim$1~--~29~minutes, from the start of the observing sequence for the 6.9~mHz, 4.6~mHz, and 2.2~mHz oscillations, respectively.

Isolating the frames, in which the oscillation is dominant, is key as it allows the spatial characteristics of the wave mode to be determined more readily. Establishing the power at each pixel within these isolated frames was performed with both Fourier and wavelet techniques. Again, each pixel of the filtered data was treated as an individual light curve with the power calculated as the absolute value of the FFT/wavelet squared. This procedure resulted in two-dimensional power plots (Figure~\ref{Fig5}) showing the spatial distribution of the power, hence revealing whether a surface or body mode is observed. One-dimensional cross cuts of these power maps produce the power profiles displayed in Figures~\ref{Fig3}b, \ref{Fig5}b and \ref{FigA1}.

The filtered time-distance diagrams reveal that the pores have coherent in-phase oscillations across their entire structure (Figure~\ref{Fig3}a). Significantly, this result highlights that the entire pore can be excited coherently, acting effectively as a monolithic magnetic flux tube. This is different from the wave behavior of the larger sunspots, which apparently do not show signatures of being excited as a single object. As such, there is no evidence in any of the pores here for the distinct `nested bowl' routinely observed in sunspot oscillations \citep{RouppeVdVoort2003}. This is probably the result of the dominant oscillations of the pores studied being globally excited, whereas in the case of sunspots, only a portion of the sunspot is excited, leading to the chevron structures in filtered time-distance cross-cuts. It is possible that the difference in physical scales between pores and sunspots are responsible for the visible wave excitation signatures, with the relatively smaller spatial size of pores allowing the underlying wave drivers to globally excite the observed waves.

The oscillatory behavior in the pore can also be contrasted with the signal in the surrounding granules. We observe significant power at periods of $\sim$300 s in the surrounding granulation, likely to result from $p$-mode leakage into the surrounding photosphere along small-scale or weak fields in the inter-granular lanes \citep{Braun1988, Li2001}. It is evident that the oscillatory behavior outside of the pores does not display the same large-scale coherence observed within the pores. Rather, the perturbations appear more random with regard to their spatial distribution and phase. The striking fact that the pores are essentially excited as a monolithic structure opens up unique and exciting avenues for studying the behavior and nature of the oscillatory phenomenon. 

It should also be noted here that it is possible that there is an enhancement of power due to small-scale reconnection at the pore boundary. However, we refute this possibility, in this case, due to a number of reasons such as the uniformity of power enhancement around the pores and the fact that the free energy is two orders of magnitude below the 1$\sigma$ error estimate (i.e., the noise estimate) of the field energy. Full details of our reasoning are given in Appendix~\ref{Reconnection}.

\subsection{Surface waves}
\label{surface}
It is found that the surface wave is the dominant oscillatory mode excited, with signatures of the mode both present in all pores examined and visible across a wide range of frequencies. Figures~\ref{Fig3} display a sample of the obtained spatial distributions of power for a pore at a range of frequencies. It may be seen from the figures that the maximum of power in the pore occurs along the interface between magnetic and non-magnetic regions, i.e., at the pore boundaries, and the power decreases to a minimum towards the center of the pore. A comparison with Figure~\ref{Fig1} reveals that the observed distribution is strikingly similar to the sketch demonstrating the key features predicted by solar MHD wave theory for surface modes. The peak power is not precisely coincident with the time-averaged pore boundary, but this is not surprising since the size and location of the pore vary as a function of time, which will naturally influence the positions of the identified boundaries. 

Figure~\ref{Fig3}a shows a time-distance cut of the filtered intensity, with the time axis limited to a 20-minute window in which the oscillation is clearly present as determined through both wavelet and EMD analyses. The corresponding power plot in panel b is constructed using this window. Note that all of the power maps are generated for times within our observing sequences when the oscillation is clearly present, as opposed to the whole duration of the series, which would act to mask out the power signal of the oscillations. It is clear in the time-distance slice of Figure~\ref{Fig3}a that there is significant wave amplitude at the boundary of the waveguide, which manifests as peaks in power in the 1-dimensional power plots. There is still some discernible intensity amplitude within the center of the pore, which is relatively weak and is not seen in the corresponding power plot. This may not be entirely unexpected as it is possible that the pore has a steep gradient in density radially, which inhibits wave power within the pore. Figure~\ref{Fig4} shows the cut used to produce the plots in Figure~\ref{Fig3} and demonstrates this phenomenon by showing the variation in light curves observed at the pore boundary and at a central location for the sample pore. It is evident in this plot that there is discernible amplitude at the center of the pore in the filtered intensity plots. However, the amplitude of the oscillations is larger towards the boundaries.

\begin{figure}[h!]
\makebox[\linewidth]{
   \includegraphics[width=0.6\linewidth]{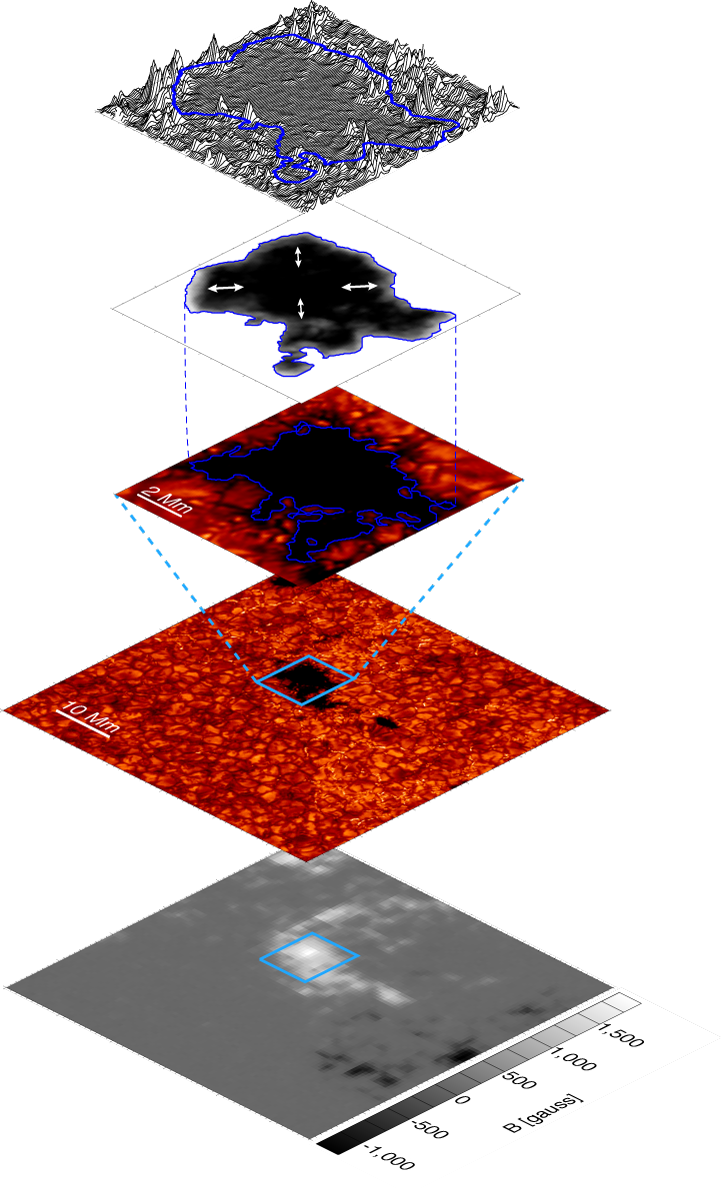}
	}
     \caption{This stack image shows the basis of our designation of a sausage surface mode in this pore. The bottom panel shows the LOS magentogram from HMI indicating the magnetic field of the pore and the sharp boundary in terms of magnetic field at the pore's edge. Above this is the full FoV ROSA G-band image showing the photospheric appearance of the pore taken on 2011 December 10. The {\textit{blue}} box indicates the expanded region shown in the three top panels. The expanded G-band image has {\textit{blue}} contours indicating the pore boundary established for that particular frame. Above this is the time-averaged pore boundary map showing the variation in boundary location during our observation sequence, where the arrows indicate the sausage mode oscillations present. The top panel is a two-dimensional power plot of the power across the pore obtained with wavelet transforms of the data filtered at a frequency of $\sim$4.6~mHz. The {\textit{blue}} contour shows the time average pore boundary location. Peaks in power at this boundary indicate a sausage mode is observed at this frequency.
}
     \label{Fig4b}
\end{figure}

In Figure~\ref{Fig4b}, a two-dimensional surface plot of the power for a surface mode oscillation at 4.6 mHz is displayed. It is clear that the power of the oscillation peaks at the boundary of the pore, with significant reduction in power across its internal region. The magnitude of the power is not homogenous around the boundary, likely due to longer-term variations in the pore boundary that will smear out the signal. In some examples, we observe relatively small power peaks within the pores. These peaks appear spatially consistent with the ingress of granules within our defined time-averaged pore boundary, during the evolution of the pores. In choosing pores for this study (see Section~\ref{obs} for our selection criteria), we also opted to choose simple structures, e.g. by neglecting pores with noticeably complex structure, e.g., light bridges. These power peaks associated with the ingress of granulation highlight the need for simple structures for the identification of the wave modes. The observed spatial distribution of power is seen for the majority of pores analyzed, occurring for many frequencies over a wide range (see Figure~\ref{FigA1} for examples). Full details of the oscillatory behavior found across all pores are given in Table~\ref{table2}.

\begin{deluxetable*}{l c c c c c c c}[hb!]
\tablecaption{Summary of observed/predicted parameters for pores studied.}           
\tablewidth{1000pt}
\tabletypesize{\scriptsize}
\tablehead{
\colhead{\textbf{Pore Property}} & \colhead{\textbf{2011 Jul 11}} & 
\colhead{\textbf{2011 Dec 09}} & \colhead{\textbf{2011 Dec 10}} & 
\colhead{\textbf{2012 Sept 30}} & \colhead{\textbf{2013 Mar 06}} & 
\colhead{\textbf{2013 Aug 17}} & \colhead{\textbf{2014 Apr 15}} 
}
\startdata
\label{table2}      
 $|B|$ (G)& 1300  & 1200 & 1200 & 820 & 850 & 990 & 1100 \\
Av. diameter (Mm) & 8.6$\pm$0.2 & 5.6$\pm$0.3 & 7.7$\pm$0.3 & 1.5$\pm$0.1 & 1.3$\pm$0.6 & 3.0$\pm$0.8 & 6.3$\pm$0.7 \\ 
$\delta$ Area (\%) & 1.61$\pm$0.49 & 1.72$\pm$0.62 & 4.01$\pm$0.88 & 2.45$\pm$0.42 & 2.80$\pm$0.21 & 2.02$\pm$0.65 & 2.22$\pm$0.34\\
Av. Intensity & 0.43$\pm$0.06 & 0.52$\pm$0.03 & 0.60$\pm$0.05 & 0.63$\pm$0.06 & 0.59$\pm$0.04 & 0.60$\pm$0.02 & 0.49$\pm$0.04 \\
$\delta$ Inten. (\%) & 1.47$\pm$0.20 & 2.16$\pm$0.34 & 2.10$\pm$0.42 & 3.15$\pm$0.19 & 1.90$\pm$0.41 & 1.09$\pm$0.39 & 2.54$\pm$0.19 \\
Observed Freq. range (mHz) & 2.0, 4.4, 9.5  & 2.1, 3.9, 7.4  & 2.2, 4.6, 6.9 & 2.8, 4.6, 11.1 & 3.6, 6.9, 11.8  & 2.5, 4.1, 7.1 & 2.9, 5.3, 10.0 \\
Surface (S) or Body (B) observed & S, S, S & S, S, S & S, S, S & B, Both, B & B, Both, B & S, S, S & S, S, S \\
Surface (S) or Body (B) predicted & S, S, S & S, S, S & S, S, S & B, S, B & B, S, B & S, S, S & S, S, S \\
Average energy estimates (kW\,m$^{-2}$)  & 43.4$\pm$11.0 & 28.8$\pm$14.7 & 6.2$\pm$1.9 & 8.6$\pm$4.8 & 8.5$\pm$3.1 & 6.3$\pm$0.8 & 41.1$\pm$10.6 \\
\enddata
\end{deluxetable*}

\subsection{Body waves}
\label{Body}
\begin{figure*}[h!]
\makebox[\linewidth]{
   \includegraphics[width=0.8\linewidth]{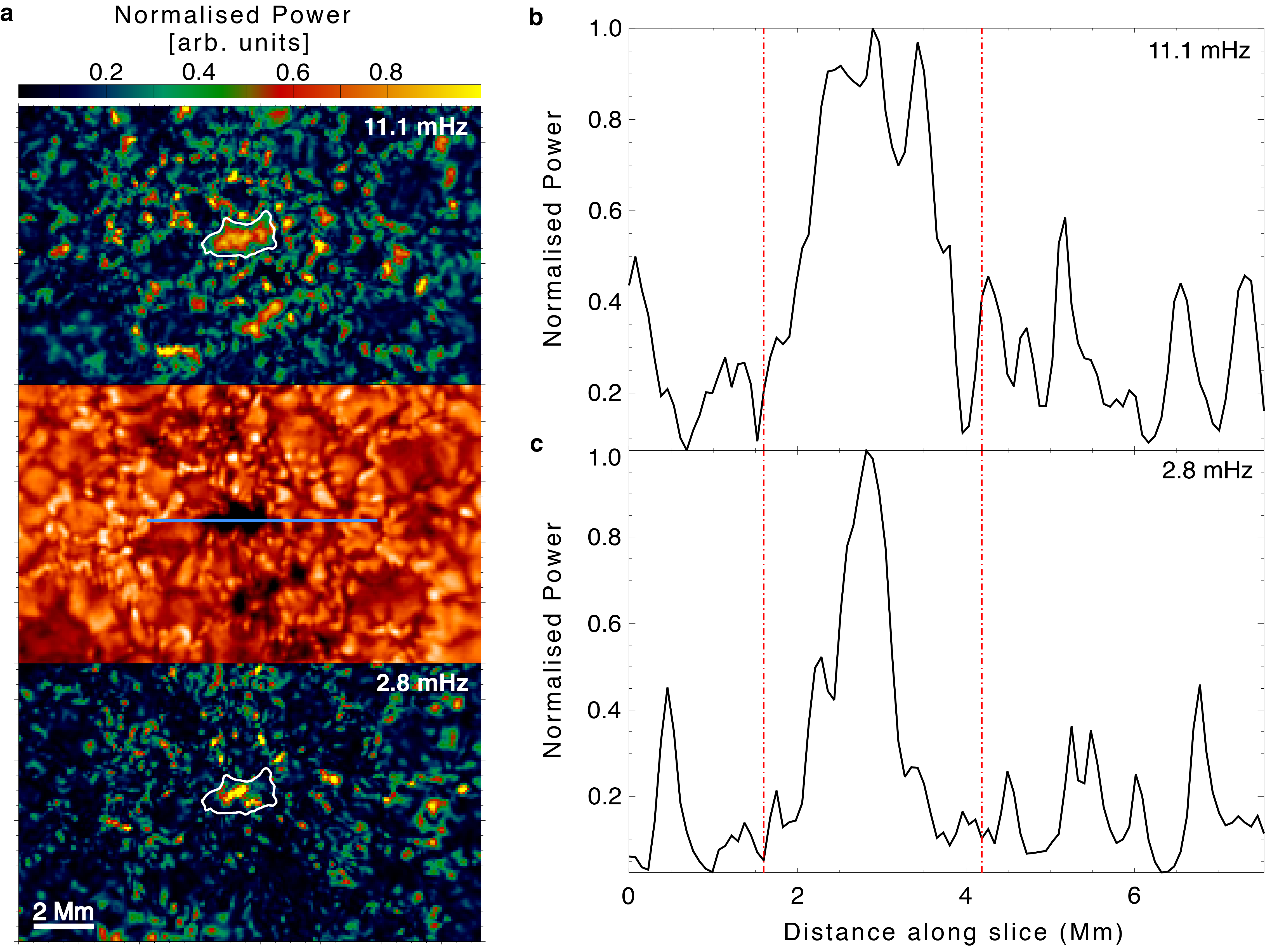}
	}
     \caption{The oscillatory signals of a near elliptical pore (central panel in {\textbf{a}}), observed on 2012 September 30, reveal concentrations of power peaking at the center of the pore. The upper and lower panels of {\textbf{a}} show the wavelet power for the pore filtered at a central frequency of 11.1~mHz and 2.8~mHz, respectively. White contours in these power maps show the time averaged boundary location for the pore. Images in {\textbf{b}} display the corresponding one dimensional cross cuts for the two power maps in {\textbf{a}}. Cuts were taken from the position marked by the {\textit{blue}} line in the central panel of {\textit{a}}. The {\textit{red}} dashed lines in {\textbf{b}} indicate the pore boundary. Again, it is evident in these cross cuts, as with the two-dimensional power plots, that power peaks within the center of the pore and decays at the boundaries, which is synonymous with the body mode.
}
     \label{Fig5}
\end{figure*}

From all the datasets examined, the distribution of power presented only two clear cases of the body mode (Figure~\ref{Fig5}), corresponding to data taken on 2012 September 30 and 2013 March 6. In both these datasets the body mode was observed unambiguously at the highest frequency ($\sim$11~mHz). The distribution of power is observed to be maximal at the center of the pore and decreases towards the boundary (and, of course, away from the boundary similar to surfaces waves). Two-dimensional power plots are displayed in Figure~\ref{Fig5}a, with the symmetry of the distribution of power clear at 11~mHz. The corresponding one-dimensional cross-cut supports this (Figure~\ref{Fig5}b), although it reveals that the observed signal is more complex than the simple schematic presented in Figure~\ref{Fig1}. For the oscillation at 2.8~mHz, there are two distinct concentrations of power within the center of the pore. This still conforms to the description of a body mode. However, it is not as clear-cut as the 11.1~mHz example. The two concentrations of power, in this instance, could be the result of a higher radial harmonic at that particular frequency, although as our examples of such a power distribution are limited, we can only speculate that this the case here. The pore in this figure is seen to possess a highly elliptical cross-sectional geometry, but this does not affect the interpretation. It is thought that potentially any cross-sectional geometry of a waveguide will allow for the existence of both body and surface modes. In particular, sausage mode oscillations of both surface and body type have been theoretically demonstrated to be supported by elliptical waveguides \citep{ErdelyiMorton2009}. 

The oscillatory signals observed within the datasets were also subject to a series of theoretical tests to predict the characteristics of the modes (see Section~\ref{predicting} for more details). Such tests involved the extraction of area and intensity information, which fed into a theoretical framework permitting the calculation of the phase velocity, which subsequently allowed us to predict the wave character. The model outputs displayed excellent agreement between the observed wave modes and those predicted by the theoretical method for the most part. Previous work \citep{Moreels2015a, Grant2015, Freij2016} employed these methods as an indirect way for detecting slow surface modes, the results of which are in agreement with what we find here.


\section{Concluding Remarks}
\label{conc}
We have demonstrated that magnetic pores are ideal features for studying the properties of MHD waves in solar magnetic wave-guides. The pores harbor a wide range of oscillatory perturbations and appear to be excited as a monolithic structure. This behavior, along with their large size, enables the radial spatial structure of the oscillatory modes to be probed. The wave behavior displays amplitude distributions that are in agreement with those predicted from theoretical models \citep{Spruit1982, EdwinRoberts1983}, although, as to be expected, the physical properties of real pores (e.g., geometry) and the surrounding atmosphere causes deviations from the simplified theory. However, the evidence presented here demonstrates in a compelling way that body and surface modes naturally exist in the Sun's atmosphere.

For the relatively small sample of pores, we found that the surface mode is more prevalent than the body mode. At present, it is unclear as to why the surface mode dominates the excited oscillatory signals. Considering all the derived parameters for the pores (Table~\ref{table2}), the only relationship between the pore parameters and modes present is the size and/or strength of LOS magnetic field. For our limited sample, we observe that for pores with diameters less than about 3~Mm and field strengths below about a kilogauss, the body modes are present. It is likely that the stronger field strength results in a sharper gradient between the pore and the quiescent environment, which allows the surface mode to be more readily supported in the pore. More research would need to be undertaken to clarify this. 

It has also been suggested that pores can support a significant amount of wave energy \citep{Grant2015}, with the potential to power the local dynamics of the lower solar atmosphere. Utilizing the theoretical framework for energy flux estimates \citep{Moreels2015b}, we suggest that at the photospheric level the surface modes transport at least twice the energy (22$\pm$10~kW\,m$^{-2}$) as the observed body modes (11$\pm$5~kW\,m$^{-2}$). This may be significant in determining which mode contributes more to localized atmospheric heating as a function of waveguide height. Again, more work needs to be done to clarify this in the context of energy deposition with height.

The ability to observe the radial spatial structuring of pores will also open up new avenues in MHD wave studies. In particular, the use of solar magneto-seismology to probe the local plasma conditions is expected to allow significant progress. We envisage that advanced models of magnetic flux concentrations embedded in a convective plasma, used in conjunction with current and further observations, will further improve our understanding of the complex physics and wave behavior that is observed within this study.


\acknowledgments{P.H.K. and R.J.M. are grateful to the Leverhulme Trust for the award of Early Career Fellowships. D.B.J. wishes to thank the UK Science and Technology Facilities Council (STFC) for the award of an Ernest Rutherford Fellowship alongside a dedicated Research Grant. D.B.J. and S.D.T.G also wish to thank Invest NI and Randox Laboratories Ltd. for the award of a Research \& Development Grant (059RDEN-1) that allowed this work to be undertaken. M.M. and F.P.K acknowledge support from the STFC Consolidated Grant to Queen's University Belfast. R.E. acknowledges the support received from the Royal Society. Armagh Observatory is funded by the Northern Ireland Department of Communities. Observations were obtained at the National Solar Observatory, operated by the association of Universities for Research in Astronomy, Inc. (AURA), under cooperative agreement with the National Science Foundation. Two datasets (2013-08-17 and 2014-04-15) were acquired within the SolarNet Project. SolarNet is a project supported by the EU-FP7 under Grant Agreement 312495. The authors wish to thank the DST staff and Dr. Gianna Cauzzi for help in acquiring data.}



\appendix

\begin{figure*}[ht!]
\makebox[\linewidth]{
   \includegraphics[width=0.73\linewidth]{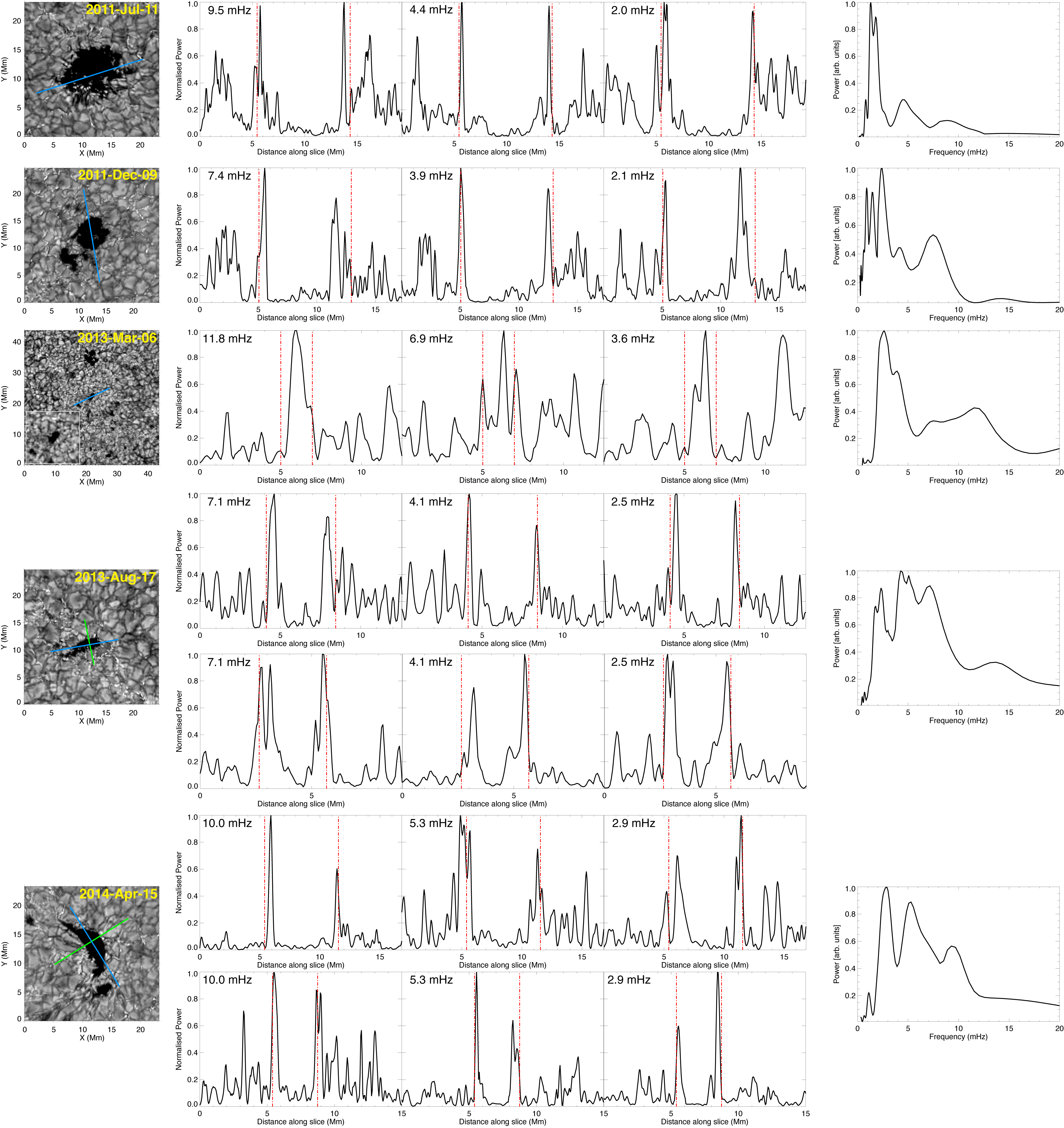}
	}
     \caption{The left column displays a sample view of five further pores analysed for this study. All images are taken with the G-band filter (4305.5~\AA) with {\textit{yellow}} labels 
in the upper right corners indicating the date of observation for the pore. A {\textit{white}} box in the 2013-Mar-06 panel shows a 10~Mm $\times$ 10~Mm zoom of the pore 
under investigation. {\textit{Blue}} lines in these images indicate the cross-cuts used to plot the 
one-dimensional power cross-cuts shown in the plots on the right of each pore. The one-dimensional power cross-cuts for each pore is acquired at the specific dominant 
oscillation frequencies observed in the data, which have been isolated with a Gaussian filter. In the top left of each cross-cut the labels indicate the central frequency employed in the 
filter, while the {\textit{red}} dashed lines indicate the pore boundary. Peaks at the boundary with a minimum in the center of the pore are a typical  feature of the surface mode, while 
peaks in the center which decay to minima at the boundaries are a characteristic of the body mode. The final two pore images have an additional {\textit{green}} line, which 
indicates an additional set of power cross-cut plots that have been included for these datasets. These cross-cuts are perpendicular to those shown from the {\textit{blue}} cross-cut. 
The upper plot panels to the right of the pores show the plots for the {\textit{blue}} line while the lower panels show the plots for the {\textit{green}} line. The bottom two pore examples 
are more elliptical in nature than the other pores and these additional plots have been included to show that the power distribution can be seen at various angles around the pore. The final plots on the far right for each dataset shows the power as a function of frequency for the pores. The highest power appears within $\sim$2\,--\,5~mHz, which would suggest that $p$-modes are responsible for exciting the waves. }
     \label{FigA1}
\end{figure*}

\section{The viability of G-band for the analysis}
\label{ViaGband}

\begin{figure*}[!h]
\makebox[\linewidth]{
   \includegraphics[width=0.75\linewidth]{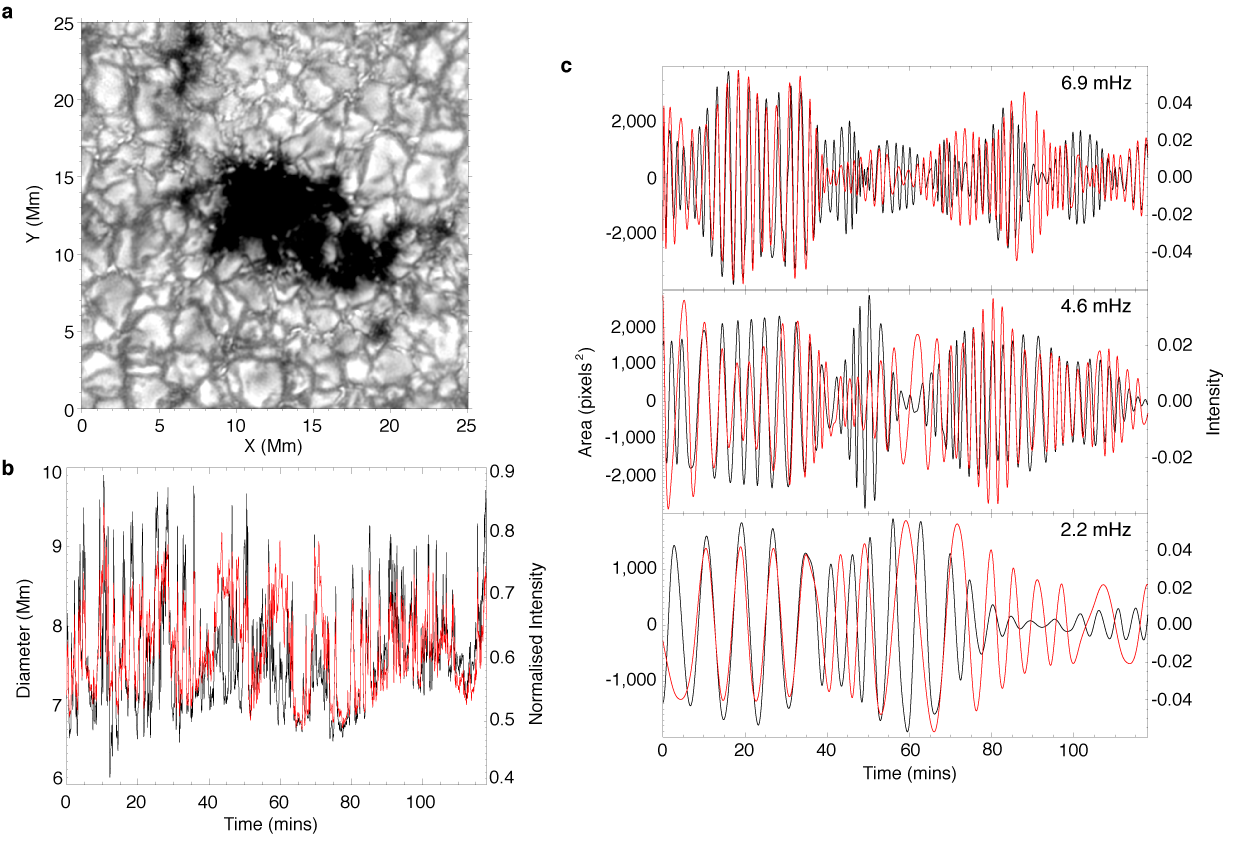}
	}
     \caption{The images presented here help to verify the decision that the G-band continuum filter is an acceptable continuum 
	bandpass for studying sausage oscillations in photospheric pores. Panel {\textbf{a}} shows a sub-field of data taken with the 4170~{\AA} continuum filter 
	for the 2011 December 10 dataset used in the paper. This is the same sub-field as depicted in Figure~\ref{Fig4}. Panel {\textbf{a}} shows the 
	visual similarities between the two filters. Panel {\textbf{b}} and {\textbf{c}} mirror plots seen in Figure~\ref{Fig2}. Panel {\textbf{b}} shows the area 
	({\textit{black}}) and intensity ({\textit{red}}) signals for the pore over the duration of the observing sequence for the 4170~{\AA} filter. The difference in intensity signals 
	between the G-band and 4170~{\AA} continuum filters is less than 2\%. Oscillations with the same period as those previously observed with the G-band filter were
	observed in the 4170~{\AA} filter area and intensity signals as well. Panel {\textbf{c}} shows the associated IMFs for these three oscillations in the 4170~{\AA} continuum 
	images. Again, {\textit{black}} represents the area and the {\textit{red}} curves show the IMFs associated with the intensity signal. A similar phase relationship 
	is observed for these signals as those observed with the G-band continuum images.
	}
     \label{FigA2}
\end{figure*}

We employed G-band in the analysis for a number of reasons. As a continuum filter, we expect that there will be a relation between the intensity of the image and temperature to some extent. This is evident when looking at MBPs in G-band images whereby the partially evacuated flux tube allows the observer to see a deeper, hotter region of the photosphere. In G-band, this increase in temperature leads to dissociation of the CH molecule and, therefore, the bright points appear brighter \citep{Steiner2001, Shelyag2004}. Thus, there is a relationship between G-band intensity and temperature, which is necessary for determining the presence of the sausage mode by analysing the variation of intensity and area signals. This is also seen when comparing the similarities of G-band to other continuum bands (Figure~{\ref{FigA2}}).

Due to the fact that more datasets within the ROSA archive employed the G-band filter to observe pores, we decided on using the G-band filter for our analysis to increase the sample size of pores and to remain consistent in our analysis of the pores we had. However, to show conclusively that G-band is an acceptable choice for our analysis, we also analysed data from another continuum filter (4170~{\AA} with a bandpass of 52~{\AA}) for one of the datasets in our G-band sample. We chose the 2011 December 10 dataset, which is shown in the main text. The 4170~{\AA} continuum data set represents the same FoV as the G-band and was operated at the same frame rate. Therefore, the cadence after post-facto image reconstruction with the 4170~{\AA} continuum is the same as the G-band (i.e., 2.112~s).

Performing precisely the same analysis as described in Section~\ref{FindOscs} of the main text on the 4170~{\AA} continuum, we find the same oscillations present in the 
data while analysing the intensity and area signals (see Figure~{\ref{FigA2}}) as we do in G-band. After filtering the data we also see the same signatures of 
the surface mode in the 4170~{\AA} continuum as we see in the G-band continuum images. This is, perhaps, unsurprising as the two filters have similar formation heights \citep{Jess2010} and the fact that similar oscillations were observed between the two filters has been observed previously for sausage modes in pores \citep{Grant2015}. Furthermore, a  study \citep{Jess2012} of the propagation characteristics of wave phenomena observed between the 4170~{\AA} continuum and the G-band continuum verified that G-band intensities can be matched to density fluctuations. It can be seen from Figure~7 of this study that the $k$-omega diagrams for both bandpasses look identical. This shows that the response of both filters is the same for input wave-like perturbations and, therefore, they are both density sensitive. The evidence presented in these studies and with our additional analysis of the 4170~{\AA} continuum data, therefore, gives further credence to the suitability of G-band in studying sausage modes in photospheric pores.

\section{Power enhancement due to reconnection}
\label{Reconnection}

\noindent In Section~\ref{Spatpow}, it is stated that reconnection may play a role in power enhancement at the pore boundaries. 
Here we will provide several reasons why we believe that this is not the case in our data, thus strengthening the case for the surface mode. The major tenets 
of our argument can be summarised as:
\begin{enumerate}
\item There are few (if any) brightenings at the pore boundary, which could be associated with reconnection phenomenon.
\item MBP motions do not support large-scale, uniform, reconnection about the pore boundary.
\item Analysis of HMI magnetic field data indicates that there is little change in the free energy in the active regions under investigation, indicating reconnection 
phenomena are fairly minor in the data.
\end{enumerate}
The first point is that, for reconnection phenomena to be responsible for the power enhancements, it would need to be on a large scale, continuously occurring around the pore boundary. As such we would expect to observe some form of intensity enhancements around the pore, corresponding to these reconnection events. Within our datasets, however, there are few such intensity enhancements (and in some instances, none at all) around the pore boundary, and definitely not at the scale that would be needed to produce the observed distribution of power. Furthermore, by filtering the data one could expect that the effects of reconnection would be diminished in the subsequent power plots, however, we still observe the power enhancement around the pore boundary.

As an extension of this point, MBPs (small-scale magnetic elements found in intergranular lanes) do not move in such a way that they could support 
the uniform reconnection pattern that would be required to produce the power plots that we find in this study. Previous work \citep{Keys2014} on the 2011 December 10 
data analysed here used a tracking algorithm to study the motion of MBPs in this particular active region. Our study came to the conclusion that the MBPs did not have a 
preferential direction of motion, that the diffusion of MBPs did not differ between active region MBPs and quiet Sun MBPs and that active region MBPs were slightly less 
dynamic that their quiet Sun counterparts. This would suggest that reconnection phenomena associated with MBPs drifting near the pore boundary, is insufficient to 
create the power enhancement associated with our surface mode observations.

\begin{figure*}[h!]
\makebox[\linewidth]{
   \includegraphics[width=1\linewidth]{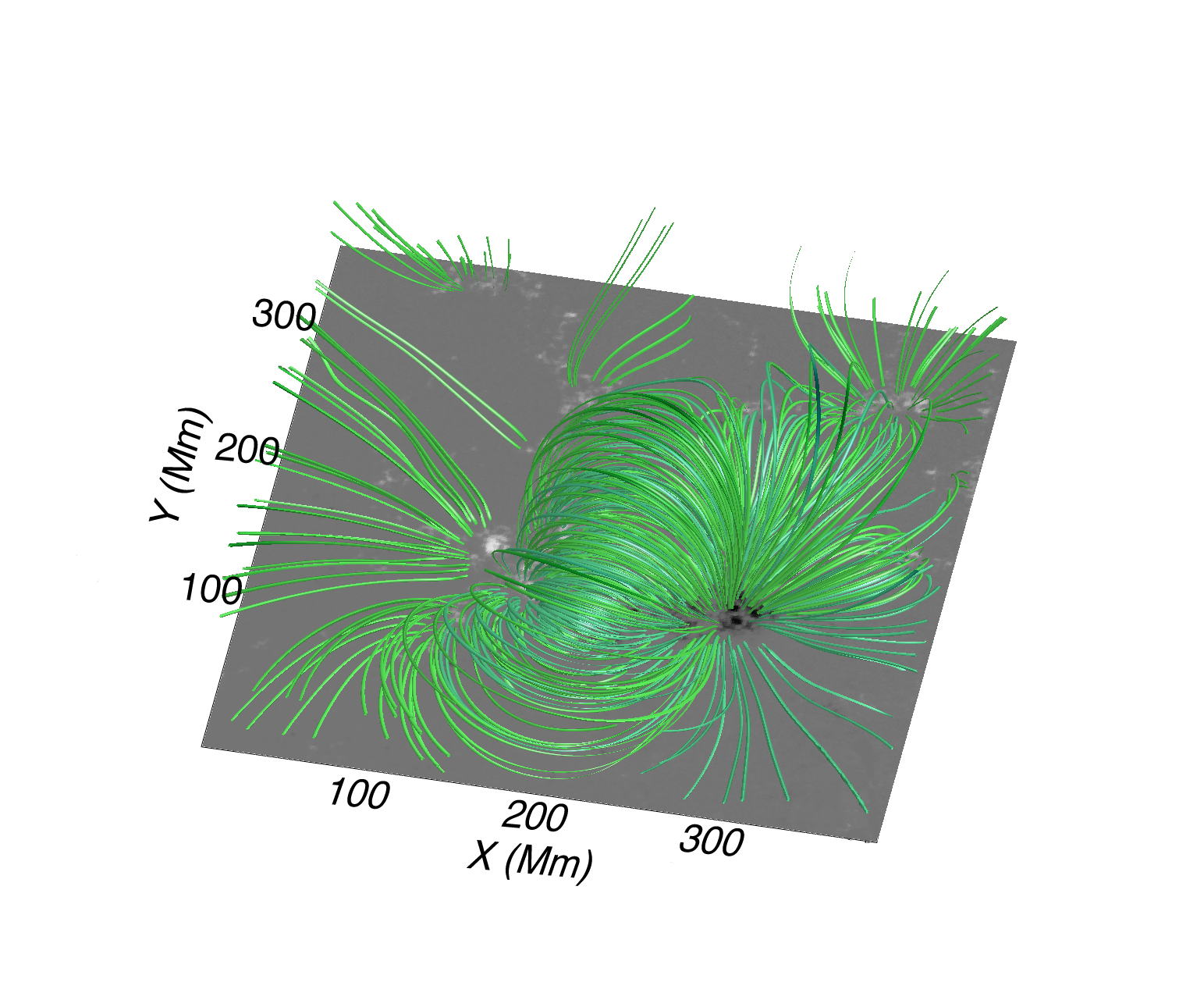}
	}
     \caption{The extrapolated magnetic field lines ({\textit{green}} lines) overlaid on the photospheric $B_z$ map from HMI. 
	This corresponds to ROSA data taken on 2011 December 10 shown in the main text, with the center of the image corresponding to the pore under analysis (see lowest image in the 
	stack in Figure~\ref{Fig4b} of the main text). The box extends to around 110~Mm above the active region. Analysing the free energies from these extrapolations suggests that 
	reconnection phenomena cannot be responsible for the power enhancement at the pore boundary for our surface mode observations in the filtered data shown in the main text. 
	Therefore, the power enhancements must be the result of wave phenomena observed in the dataset.}
     \label{FigA3}
\end{figure*}

We can quantify further whether reconnection phenomena are present in the data by employing a non-linear force-free field extrapolation code \citep{Wiegelmann2008} on vector magnetograms obtained with HMI. An example of such an extrapolation can be seen in Figure~\ref{FigA3} using the 2011 December 10 data. Analysing the active region as a whole, i.e., including magnetic regions outside our ROSA FoV, we estimate that the free energy (the difference between the non-potential and potential volume magnetic field energies) decreases by 1.6$\times$10$^{27}$~erg over the course of the dataset. However, the estimated 1$\sigma$ noise threshold is 1.4$\times$10$^{29}$~erg, so the predicted change in free energy is nearly two orders of magnitude below the error estimate of the field energy. The errors associated with the magnetic free energy have been propagated in accordance with \citet{GeorgoulisLaBonte2007}, who calculated the relative magnetic helicities and free energies with respect to a potential-field reference, with a detailed analysis of the error calculations presented in Appendix B of \citet{Georgoulis2012}. With no macroscopic signatures of (micro)flaring around the pores and minuscule change in the active region's free energy (which is embedded within the noise limit of the extrapolation code), this would suggest that any reconnection phenomena is exceptionally weak, and not enough to produce the power enhancement in our filtered power plots. Therefore, due to these various reasons, we suggest that the power enhancement we observe is the result of wave phenomena present within the data.

\begin{thebibliography}{}
\bibitem[Aschwanden et al.(2004)]{Aschwanden2004} Aschwanden, M.~J., Nakariakov, V.~M., \& Melnikov, V.~F.\ 2004, \apj, 600, 458
\bibitem[Banerjee et al.(2007)]{Banerjee2007} Banerjee, D., Erd{\'e}lyi, R., Oliver, R., \& O'Shea, E.\ 2007, \solphys, 246, 3  
\bibitem[Beck et al.(2007)]{Beck2007} Beck, C., Bellot Rubio, L.~R., Schlichenmaier, R., \& S{\"{u}}tterlin, P.\ 2007, \aap, 472, 607 
\bibitem[Bogdan \& Judge(2006)]{BogdanJudge2006} Bogdan, T.~J., \& Judge, P.~G.\ 2006, Philosophical Transactions of the Royal Society of London Series A, 364, 313 
\bibitem[Braun et al.(1988)]{Braun1988} Braun, D.~C., Duvall, T.~L., Jr., \& Labonte, B.~J.\ 1988, \apj, 335, 1015
\bibitem[De Moortel \& Nakariakov(2012)]{DeMoortelNakariakov2012} De Moortel, I., \& Nakariakov, V.~M.\ 2012, Philosophical Transactions of the Royal Society of London Series A, 370, 3193  
\bibitem[De Moortel(2009)]{DeMoortel2009} De Moortel, I.\ 2009, \ssr, 149, 65 
\bibitem[De Pontieu et al.(2004)]{DePontieu2004} De Pontieu, B., Erd{\'e}lyi, R., \& James, S.~P.\ 2004, \nat, 430, 536
\bibitem[De Pontieu et al.(2005)]{DePontieu2005} De Pontieu, B., Erd{\'e}lyi, R., \& De Moortel, I.\ 2005, \apjl, 624, L61  
\bibitem[Defouw(1976)]{Defouw1976} Defouw, R.~J.\ 1976, \apj, 209, 266
\bibitem[D{\'{\i}}az \& Roberts(2006)]{DiazRoberts2006} D{\'{\i}}az, A.~J., \& Roberts, B.\ 2006, \aap, 458, 975  
\bibitem[Dorotovi{\v{c}} et al.(2008)]{Dorotovic2008} Dorotovi{\v{c}}, I., Erd{\'{e}}lyi, R., \& Karlovsk{\'{y}}, V.\ 2008, Waves \& Oscillations in the Solar Atmosphere: Heating and Magneto-Seismology, 247, 351 
\bibitem[Dorotovi{\v{c}} et al.(2014)]{Dorotovic2014} Dorotovi{\v{c}}, I., Erd{\'{e}}lyi, R., Freij, N., Karlovsk{\'{y}}, V., \& M{\'{a}}rquez, I.\ 2014, \aap, 563, A12 
\bibitem[Dorotovi{\v c} et al.(2016)]{Dorotovic2016} Dorotovi{\v{c}}, I., Rybansk{\'{y}}, M., Sobotka, M., et al.\ 2016, Coimbra Solar Physics Meeting: Ground-based Solar Observations in the Space Instrumentation Era, 504, 37 
\bibitem[Edwin \& Roberts(1983)]{EdwinRoberts1983} Edwin, P.~M., \& Roberts, B.\ 1983, \solphys, 88, 179 
\bibitem[Erd{\'{e}}lyi \& Morton(2009)]{ErdelyiMorton2009} Erd{\'{e}}lyi, R., \& Morton, R.~J.\ 2009, \aap, 494, 295
\bibitem[Erd{\'{e}}lyi \& Fedun(2010)]{ErdelyiFedun2010} Erd{\'{e}}lyi, R., \& Fedun, V.\ 2010, \solphys, 263, 63  
\bibitem[Freij et al.(2014)]{Freij2014} Freij, N., Scullion, E.~M., Nelson, C.~J., et al.\ 2014, \apj, 791, 61 
\bibitem[Freij et al.(2016)]{Freij2016} Freij, N., Dorotovi{\v{c}}, I., Morton, R.~J., et al.\ 2016, \apj, 817, 44 
\bibitem[Grant et al.(2015)]{Grant2015} Grant, S.~D.~T., Jess, D.~B., Moreels, M.~G., et al.\ 2015, \apj, 806, 132 
\bibitem[Grinstead et al. (2004)]{Grinstead2004} Grinsted, A., Moore, J.~C., \& Jevrejeva, S. \ 2004, Nonlinear Processes in Geophysics, European Geosciences Union (EGU), 11 (5/6), pp. 561-566.
\bibitem[Georgoulis \& LaBonte(2007)]{GeorgoulisLaBonte2007} Georgoulis, M.~K., \& LaBonte, B.~J.\ 2007, \apj, 671, 1034 
\bibitem[Georgoulis et al.(2012)]{Georgoulis2012} Georgoulis, M.~K., Tziotziou, K., \& Raouafi, N.-E.\ 2012, \apj, 759, 1 
\bibitem[Goedbloed \& Poedts(2004)]{GoedbloedPoedts2004} Goedbloed, J.~P.~H., \& Poedts, S.\ 2004, ``Principles of Magnetohydrodynamics, by J.P.H.~Goedbloed and S.~Poedts.~ISBN 0521626072.
\bibitem[Huang et al.(1998)]{Huang1998} Huang, N.~E., Shen, Z., Long, S.~R., et al.\ 1998, Proceedings of the Royal Society of London Series A, 454, 903 
\bibitem[Huang \& Wu(2008)]{Huang2008} Huang, N.~E., \& Wu, Z.\ 2008, Reviews of Geophysics, 46, RG2006 
\bibitem[Jess et al.(2010)]{Jess2010} Jess, D.~B., Mathioudakis, M., Christian, D.~J., et al.\ 2010, \solphys, 261, 363
\bibitem[Jess et al.(2012)]{Jess2012} Jess, D.~B., Shelyag, S., Mathioudakis, M., et al.\ 2012, \apj, 746, 183 
\bibitem[Jess et al.(2015)]{Jess2015} Jess, D.~B., Morton, R.~J., Verth, G., et al.\ 2015, \ssr, 190, 103 
\bibitem[Keys et al.(2014)]{Keys2014} Keys, P.~H., Mathioudakis, M., Jess, D.~B., Mackay, D.~H., \& Keenan, F.~P.\ 2014, \aap, 566, A99 
\bibitem[Li et al.(2001)]{Li2001} Li, W., Ai, G., \& Wang, H.\ 2001, Publications of the Beijing Astronomical Observatory, 37, 15 
\bibitem[Ludwig et al.(2009)]{Ludwig2009} Ludwig, H.-G., Samadi, R., Steffen, M., et al.\ 2009, \aap, 506, 167 
\bibitem[Luna-Cardozo et al.(2012)]{LunaCardoza2012} Luna-Cardozo, M., Verth, G., \& Erd{\'{e}}lyi, R.\ 2012, \apj, 748, 110  
\bibitem[Maltby et al.(1986)]{Maltby1986} Maltby, P., Avrett, E.~H., Carlsson, M., et al.\ 1986, \apj, 306, 284 
\bibitem[Mathioudakis et al.(2013)]{Mathioudakis2013} Mathioudakis, M., Jess, D.~B., \& Erd{\'e}lyi, R.\ 2013, \ssr, 175, 1 
\bibitem[Moreels et al.(2013)]{Moreels2013} Moreels, M.~G., Goossens, M., \& Van Doorsselaere, T.\ 2013, \aap, 555, A75 
\bibitem[Moreels \& Van Doorsselaere(2013)]{MoreelsVD2013} Moreels, M.~G., \& Van Doorsselaere, T.\ 2013, \aap, 551, A137
\bibitem[Moreels et al.(2015a)]{Moreels2015a} Moreels, M.~G., Freij, N., Erd{\'{e}}lyi, R., Van Doorsselaere, T., \& Verth, G.\ 2015, \aap, 579, A73 
\bibitem[Moreels et al.(2015b)]{Moreels2015b} Moreels, M.~G., Van Doorsselaere, T., Grant, S.~D.~T., Jess, D.~B., \& Goossens, M.\ 2015, \aap, 578, A60  
\bibitem[Morton et al.(2011)]{Morton2011} Morton, R.~J., Erd{\'{e}}lyi, R., Jess, D.~B., \& Mathioudakis, M.\ 2011, \apjl, 729, L18 
\bibitem[Morton et al.(2012)]{Morton2012} Morton, R.~J., Verth, G., Jess, D.~B., et al.\ 2012, Nature Communications, 3, 1315
\bibitem[Morton et al.(2015)]{Morton2015} Morton, R.~J., Tomczyk, S., \& Pinto, R.\ 2015, Nature Communications, 6, 7813 
\bibitem[Nakariakov \& Verwichte(2005)]{NakariakovVerwichte2005} Nakariakov, V.~M., \& Verwichte, E.\ 2005, Living Reviews in Solar Physics, 2, 3 
\bibitem[Pesnell et al.(2012)]{Pesnell2012} Pesnell, W.~D., Thompson, B.~J., \& Chamberlin, P.~C.\ 2012, \solphys, 275, 3 
\bibitem[Rae \& Roberts(1983)]{RaeRoberts1983} Rae, I.~C., \& Roberts, B.\ 1983, \solphys, 84, 99 
\bibitem[Rimmele(2004)]{Rimmele2004} Rimmele, T.~R.\ 2004, \procspie, 5490, 34 
\bibitem[Roberts \& Webb(1978)]{RobertsWebb1978} Roberts, B., \& Webb, A.~R.\ 1978, \solphys, 56, 5 
\bibitem[Rouppe van der Voort et al.(2003)]{RouppeVdVoort2003} Rouppe van der Voort, L.~H.~M., Rutten, R.~J., S{\"u}tterlin, P., Sloover, P.~J., \& Krijger, J.~M.\ 2003, \aap, 403, 277 
\bibitem[Sakurai et al.(1991)]{Sakurai1991} Sakurai, T., Goossens, M., \& Hollweg, J.~V.\ 1991, \solphys, 133, 247 
\bibitem[Schou et al.(2012)]{Shou2012} Schou, J., Borrero, J.~M., Norton, A.~A., et al.\ 2012, \solphys, 275, 327 
\bibitem[Shelyag et al.(2004)]{Shelyag2004} Shelyag, S., Sch{\"u}ssler, M., Solanki, S.~K., Berdyugina, S.~V., \& V{\"o}gler, A.\ 2004, \aap, 427, 335 
\bibitem[Sobotka(2003)]{Sobotka2003} Sobotka, M.\ 2003, Astronomische Nachrichten, 324, 369 
\bibitem[Spruit(1982)]{Spruit1982} Spruit, H.~C.\ 1982, \solphys, 75, 3  
\bibitem[Steiner et al.(2001)]{Steiner2001} Steiner, O., Hauschildt, P.~H., \& Bruls, J.\ 2001, \aap, 372, L13 
\bibitem[S{\"{u}}tterlin et al.(1996)]{Sutterlin1996} S{\"{u}}tterlin, P., Schr{\"{o}}ter, E.~H., \& Muglach, K.\ 1996, \solphys, 164, 311 
\bibitem[Terradas et al.(2004)]{Terradas2004} Terradas, J., Oliver, R., \& Ballester, J.~L.\ 2004, \apj, 614, 435 
\bibitem[Torrence \& Compo(1998)]{TorrCompo1998} Torrence, C., \& Compo, G.~P.\ 1998, Bulletin of the American Meteorological Society, 79, 61 
\bibitem[Vernazza et al.(1981)]{VAL1981} Vernazza, J.~E., Avrett, E.~H., \& Loeser, R.\ 1981, \apjs, 45, 635 
\bibitem[Verma \& Denker(2014)]{VermDen2014} Verma, M., \& Denker, C.\ 2014, \aap, 563, A112 
\bibitem[Wang(2011)]{Wang2011} Wang, T.\ 2011, \ssr, 158, 397 
\bibitem[Wiegelmann(2008)]{Wiegelmann2008} Wiegelmann, T.\ 2008, Journal of Geophysical Research (Space Physics), 113, A03S02 
\bibitem[W{\"{o}}ger et al.(2008)]{Woger2008} W{\"{o}}ger, F., von der L{\"{u}}he, O., \& Reardon, K.\ 2008, \aap, 488, 375 
\bibitem[Yu et al.(2017a)]{Yu2017a} Yu, D.~J., Van Doorsselaere, T., \& Goossens, M.\ 2017, \aap, 602, A108 
\bibitem[Yu et al.(2017b)]{Yu2017b} Yu, D.~J., Van Doorsselaere, T., \& Goossens, M.\ 2017, \apj, 850, 44 
\bibitem[Zhugzhda et al.(2000)]{Zhugzhda2000} Zhugzhda, Y.~D., Balthasar, H., \& Staude, J.\ 2000, \aap, 355, 347 
\end{thebibliography}
\end{document}